\documentclass[doublecol]{epl2} 

\usepackage{amsmath,amssymb}
\usepackage{graphicx,dsfont}

\DeclareGraphicsExtensions{.eps,.ps}

\begin{document}

\title{Triggering up states in all-to-all coupled neurons
\\\mbox{}\hfill  \normalsize\rm epl Europhysics Letters, in print (2010)
}

\author{H.-V.V. Ngo\inst{1,2}, J. K\"ohler\inst{3}, J. Mayer\inst{3},
J.C. Claussen\inst{1,2}\thanks{Both senior authors contributed equally \protect\\[-5.8ex]}
\addtocounter{footnote}{-1}
\and H.G. Schuster\inst{3}\thanks{}} 
\shortauthor{H.-V. V. Ngo \etal}
\institute{                    
  \inst{1} Institute for Neuro- and Bioinformatics - University of L\"ubeck, 23538 L\"ubeck, Germany \\
  \inst{2} Graduate School for Computing in Medicine and Life Science - University of L\"ubeck, Germany \\
		\inst{3} Institute for Theoretical Physics and Astrophysics - University of Kiel, 24098 Kiel, Germany
}


\date{October 30, 2009}

\abstract{Slow-wave sleep in mammalians is characterized by a change of large-scale cortical activity currently paraphrased as cortical  Up/Down states. A recent experiment demonstrated a bistable collective behaviour in ferret slices, with the remarkable property that the Up states can be switched on and off with pulses, or excitations, of same polarity; whereby the effect of the second pulse significantly depends on the time interval between the pulses. Here we present a simple time discrete model of a neural network that exhibits this type of behaviour, as well as quantitatively reproduces the time-dependence found in the experiments.}

\maketitle

\section{Introduction}
Collective oscillations in networks composed of coupled elements have continuously attracted great interest regardless of the scientific field \cite{wiesenfeld1994,kiss2004,kuramoto1984,strogatz2003,pikovsky2001}. In neural systems, oscillations have been studied both experimentally and theoretically for decades \cite{singer1989,schuster1990} but a concise understanding of the neural dynamics still remains a continuous challenge.

A remarkable change of its oscillatory behaviour is displayed by the mammalian brain at the onset of sleep. The frequency of collective cortical activity, as measured by scalp electroencephalography (EEG), drastically decays during sleep onset. With further deepening of sleep a significant increase of power in the lower frequency bands delta (1-4 Hz) and theta (4-8 Hz) is observed \cite{rechtschaffen1986}.

In contrast to these clearly observed differences between sleep and wake EEG, only hypotheses for the role of these dynamical changes exist. During slow wave sleep, a sleep stage named after its oscillations in the 1 Hz region and below, cortical regions largely undergo alternations between states of high activity by synchronous firing (Up states) and stages of relatively long quiescence (Down states) \cite{destexhe1999,moelle2002}. These slow oscillations manifest an interesting phenomenon on its own, e.g. by forming travelling waves \cite{richardson2005,coombes2005,koehler2008}.

Recent studies have shown that slow oscillations during sleep may possess a promoting role in neural plasticity and learning, e.g. an enhancement of the consolidation of new memories \cite{stickgold2005}. For instance, improvement of declarative memory performance was observed after transcranial stimulation during sleep \cite{marshall2006}. A natural extension of this finding is to investigate the manipulation of slow oscillations by external stimulation \cite{marshall2006,massimini2007,bergmann2008}. This could provide an access route to influence or even control the sleep state itself, as slow waves not only were observed experimentally to group so-called spindle oscillations \cite{moelle2002}, but dynamically have the potential to entrain thalamic networks into collective sleep spindle oscillations \cite{mayer2007}. On the other hand, manipulation of slow oscillations exhibits a starting point to advance the transition of different sleep stages and in the long term even to facilitate the onset of sleep.

In recent in vitro experiments on prefrontal and occipital cortex slices of ferrets, Shu \etal \cite{shu2003} demonstrated a neural system showing collective oscillations that can be forced into an Up state by injecting a positive electrical pulse. This state could afterwards be terminated by a second pulse of identical polarity, but with a specific dependency on pulse intensity and delay to the first one \cite[fig. 1a,b]{shu2003}. This type of response on such a stimulation protocol offers a very intriguing way to control neural systems. In this paper we specifically aim to reproduce and explain the experimental results of Shu et al. by a simple but sufficiently accurate model.

The paper is organized as follows. First we introduce our model, where we demonstrate the mechanism underlying its basic behaviour by means of a single neuron. On the basis of this model we explain the experimental results of \cite{shu2003} by applying their stimulation protocol and reproduce the central response diagram relating interstimulus interval and Up state duration. We continue with an examination of the dynamics and properties of a noisy network and in particular focus on the importance of coupling. Finally, we provide a mean field analysis of the network to investigate its stability properties, after which we conclude with a brief summary and outlook.

\section{The model} 
Our model is based on the central hypothesis that slow oscillations are caused by an interplay between recurrent excitation and an inhibition controlled by slow adaptive currents, in particular a Na$^+$ dependent K$^+$ current \cite{sanchezvives2000,compte2003}. The qualitative sketch of the dynamics is as follows: During the inactive state, neurons receive synaptic input from ongoing spontaneous activity, which occasionally excites enough postsynaptic neurons and results in synchronous firing of greater regimes of cells, forming the beginning of an Up state. This state is then maintained by the recurrent excitation between all participating cells. During the burst state, inhibitory currents build up and slowly decrease the neuronal excitability, eventually commencing the inactive phase. As the majority of neurons is in their ground state, the adaptive currents recover from the integration back to their initial point, and finally allow the cycle to start again.

We introduce our model by looking at a single neuron. As the focus lies on the dynamics of the network and, above all, on the slow oscillations that operate on a much larger time scale then a single spike, we restrict our study to an integrate and fire model with McCulloch \& Pitts neurons \cite{mcculloch1943}, i.e. the neuron is reduced to two states, being either active or in its ground state, corresponding to the values 1 and 0, respectively. The activity of this binary neuron is modelled by a Heaviside step function $\Theta$, eq.\ (\ref{eq1a}). Consequently, we also neglect the dead-time of a neuron after a spike is produced. The complete model is then defined by a time-discrete map consisting of the following equations:
\begin{subequations}
 \begin{align}
  x^{t+1} & = \Theta[I-d_f-\vartheta^t] \label{eq1a} \\
	\mu^{t+1} & = \lambda_\mu\, \mu^t+g\, x^t \label{eq1b} \\
	\vartheta^{t+1} & = \lambda_\vartheta\,\vartheta^t+h\,\Theta[\mu^t-d_b]~. \label{eq1c}
 \end{align}
\end{subequations}

By construction, each neuron receives the same input $I$, which we assume to be a constant driving force, in order to explain the dynamics of our model. If we first consider the input $I$ to be slightly larger than both negative terms $d_f$ and $\vartheta^t$, where $d_f$ is the constant firing threshold and becomes adaptive in combination with $\vartheta$, the argument of the Heaviside step function is positive and hence the neuron is active, i.e.\ $x^t=1$.

Besides a leakage term in eq.\ (\ref{eq1b}) the variable $\mu$ holds a dependence on the activity of the neuron. Whenever the neuron is active, i.e.\ equal to 1, $\mu$ is increased by the value $g$. An Up state of prolonged neuronal firing thus results in a steady raise of $\mu$. This integration process eventually activates the second Heaviside step function in eq.\ (\ref{eq1c}) when the value of $\mu$ exceeds the burst threshold $d_b$.

By triggering the second Heaviside function, $\vartheta$ is immediately increased by $h$ which also effects the adaptive threshold. As this threshold is now larger than input $I$ the argument of the Heaviside step function in eq.\ (\ref{eq1a}) is no longer positive and the neuron is forced into its ground state, i.e.\ a transition from an Up to a Down state occurs.

\begin{figure}[t]
 \centering
	\includegraphics[scale=.4]{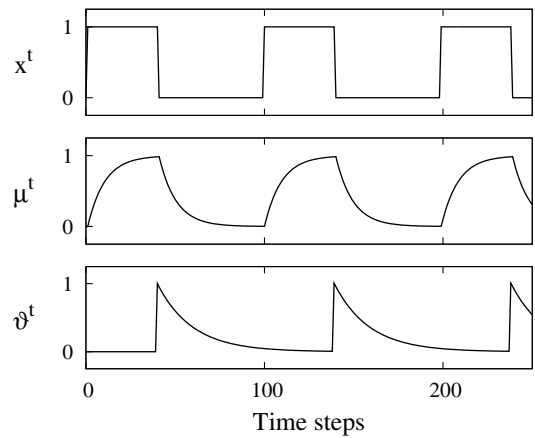}
	\caption{Dynamics of a single neuron over 250 time steps. The neuron activity oscillates between an active and quiescent phase (top). $\mu$ is responsible for the duration of an Up state, by reaching $d_b=0.98$ it triggers the inhibition and induces a Down state. Hence the interval of inactivity is determined by the decay of $\vartheta$. The remaining parameters are as follows: $I=0.25$, $d_f=0.2$, $\lambda_\mu=0.9$,	$\lambda_\vartheta=0.95$, $g=0.1$ and $h=1$.}
	\label{fig1}
\end{figure}

During this time of quiescence $\vartheta$ decays due to its leakage term. If this decay has progressed far enough, the input $I$ is eventually larger then the sum of $d_b$ and $\vartheta^t$ and the neuron moves into the Up state again, starting a new cycle. Fig.\ \ref{fig1} shows these dynamics for a few cycles. We can clearly distinguish the integration, inhibition and recovery processes responsible for the generation of slow oscillations as proposed by \cite{sanchezvives2000}. In a subsequent section we will present some analytical mean field calculations of the model, where we will have a more detailed look on the existing fixed points and their stability.

The great advantage, besides the small computational requirements to evaluate discrete maps, is that a neuron consists of only three equations, making it very easy to perform simulations. To be more specific, throughout the paper we will choose the parameters in such a way, that Up and Down states last about 50 time steps. The model parameters for the noisy network, see beneath, are, unless mentioned otherwise: $d_f=0.2$, $d_b=0.98$, $\sigma=0.2$, $C=1$, $\lambda_\mu=0.9$, $\lambda_\vartheta=0.96$, $g=0.1$ and $h=2$. 

\begin{figure}[t]
 \centering
	\includegraphics[scale=.4]{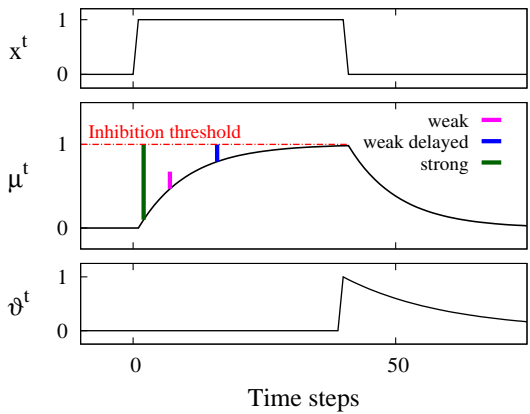}
	\caption{(Colour online) External stimuli moves neuron towards inhibition threshold. The effect of an external stimulus depends on the progress of $\mu$'s integration process, i.e. how close $\mu$ is to the inhibition threshold $d_b$ (red dashed line). Early injection of a weak pulse (magenta) will not terminate an Up state, unless the delay to the onset is larger (blue). Consequently pulses of high intensities are more likely to push $\mu$ above $d_b$ (green).}
	\label{fig2}
\end{figure}

\section{Influence of external stimuli}
Using this simple model as a basis we can give an explanation to the experimental results from Shu et al., i.e.\ the dependence of an Up state termination on the stimulus intensity and the interstimulus interval. The central role of the underlying mechanism is played by the variable $\mu$. Its dynamics reflect the build-up of inhibitory currents during an Up state as proposed by Sanchez-Vives and triggers the transition to a Down phase.

The external stimulus basically moves the variable $\mu$ closer to the inhibition threshold $d_b$. Fig.\ \ref{fig2} demonstrates this issue. Regarding the time dependence we can see that an early applied weak pulse will not be able to terminate an Up state, it will rather shorten the Up state duration by increasing $\mu$. But if the pulse injection after the onset of a burst is delayed, the steady increase of $\mu$ by the integration process has progressed far enough, such that the weak pulse is eventually strong enough to push $\mu$ above $d_b$ and trigger the inhibition mechanism by $\vartheta$. On the other hand it is obvious that with increasing pulse intensity the minimal delay necessary to terminate an Up state decays, until it is completely negligible for strong intensities, which explains the second dependency found in the experiment.

\begin{figure}[t]
 \centering
	\includegraphics[scale=.4]{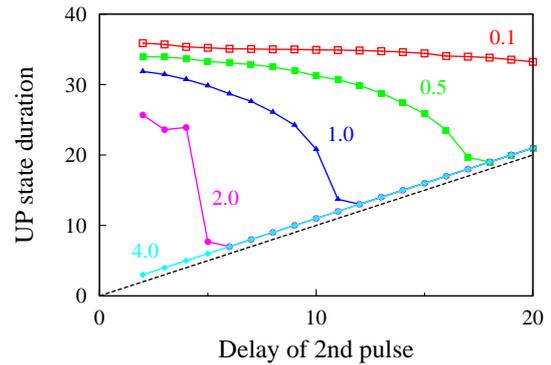}
	\caption{(Colour online) Up state duration vs. double pulse interstimulus interval. The results of a double stimulation with different intensities and interstimulus intervals show the same qualitative dependencies as found in the experiment by \cite{shu2003}, i.e.\ with increasing intensity the need of a delay vanishes. The different colours indicate different intensities.}
	\label{fig3}
\end{figure}

The application of the same stimulation protocol from \cite{shu2003}, i.e.\ different combinations of intensity and delay, on our model is summarized in fig.\ \ref{fig3}. It is clear that our results give a very satisfactory qualitative reproduction of the experimental results.

\section{A noisy network}
Next we investigated an extension of our model to a network of $N=10\,000$ neurons, where we only consider an excitatory all-to-all coupling among the neurons. Despite the fact that experimental as well as theoretical findings hint towards a balance between excitation and inhibition to sustain Up states in cortical networks \cite{durstewitz2000,wang2001}, inhibitory synaptic connection were neglected as excitatory interaction between the neurons on its own is sufficient for our investigations.

To construct the network out of eqs.\ (\ref{eq1a}-\ref{eq1c}), we removed the constant driving force $I$ and replaced it by a sum over all neurons in the network multiplied it by a normalized coupling strength $C$ and furthermore added a local Gaussian noise term $\xi$ with zero mean and intensity (or variance) $\sigma$ due to the fact that every neuron in general receives input from multiple sources of noise, e.g.\ synaptic noise and channel noise \cite{gerstner2002}. The modified system of equations reads:
\begin{subequations}
 \begin{align}
  x_i^{t+1} & = \Theta\Biggl[\frac{C}{N}\sum^N_{j=1}x^t_j+\xi_i^t-d_f-\vartheta_i^t\Biggr] \label{eq2a} \\
		\mu_i^{t+1} & = \lambda_\mu\, \mu_i^t+g\, x^t_i \label{eq2b} \\
		\vartheta_i^{t+1} & = \lambda_\vartheta\, \vartheta^t_i+h\,\Theta[\mu_i^t-d_b]~. \label{eq2c}
 \end{align}
	\label{eq2}
\end{subequations}%
Taking the Gaussian noise fluctuations into account, all following simulation results were averaged over 1000 realizations.

\begin{figure}[t]
 \centering
	\includegraphics[scale=.4]{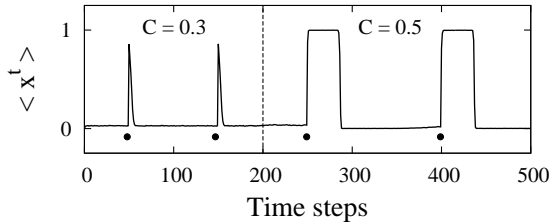}
	\caption{Role of coupling within the noisy network ($\sigma=0.1$). The noise driven network needs excitatory connections between neurons with a certain strength to generate slow oscillations. Black circles indicate external pulse injections.}
	\label{fig4}
\end{figure}

The first question to clarify is whether the network can generate slow oscillations. If we set the coupling $C=0$ and do not consider any external manipulation, the Gaussian noise $\xi$ replaces the constant input $I$ from the previous section and acts as an intrinsic driving force. Setting the variance $\sigma$ close to the firing threshold $d_f$ ensures a repetitive activation of the neurons. Due to noise fluctuations the probability of synchronous firing, which in this case means that more than 75\% of the population are simultaneously active, of a large neural population is very small. As is the case that the inhibition mechanism is activated through continuous firing of the neuron. This demonstrates the necessity of an interaction between neurons.

By increasing the coupling strength $C$ we enable an excitation of other neurons that leads to synchronous firing. More importantly, sufficient coupling strength can compensate for the noise fluctuations and the high activity within the network will be sustained. That is, this sketched process corresponds to an Up state onset. Although the model is quite simple and neglects the spiking property of single neurons, the need of excitatory connections between neurons qualitatively resembles the recurrent excitation mentioned above, which is responsible for maintaining the Up state. Persistant firing of a single neuron also causes a steady increase of $\mu$ that eventually triggers the knockout through $\vartheta$, resulting in a stable alternation between Up and Down phase.

A far more interesting question to address is to what extent it is possible to self-generate and to induce Up states in our network. Beginning with a high coupling strength and slowly decreasing it, the ability to self-generate slow oscillations is lost after passing a critical point. The network shows activity but homogeneously distributed and not grouped together. But for a certain range it is still possible to induce a complete Up state just by applying a single pulse as fig.\ \ref{fig4} shows. We only studied the stimulation of all neurons, but depending on the present coupling it is possible to apply the pulse only on a subpopulation.

From the left part of fig.\ \ref{fig4} we can conclude another important aspect, namely that in this range of coupling strength, there is no dependence on the pulse intensity. Injecting a higher intensity on all neurons would only result in a very short excitation of the neurons, as the coupling strength is too weak to maintain an Up state.

\begin{figure}[t]
 \centering
	\includegraphics[scale=.4]{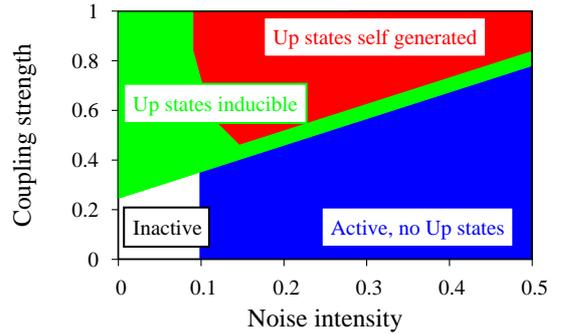}
	\caption{(Colour online) Phase portrait of noisy network. Variation of coupling strength and noise intensity (i.e. variance 		$\sigma$) reveals three firing states of the network. Besides the phase in which Up states formed of at
	  least 75\% active neurons of the network population are intrinsically generated (red), the other two represent
	  phases where Up states may be induced by an external (periodic) signal or the networks shows random distributed
	  activity with no Up states.}
	\label{fig5}
\end{figure}

Further reduction of $C$ finally yields a state where, only due to sufficient noise intensity, the network is again active with a low amplitude homogeneously distributed over the neurons. Fig.\ \ref{fig5} shows a phase portrait of these three states that have just been outlined. There is in fact a fourth phase, in which the network shows no activity at all, but that does not play an important role, as the brain itself does not show a state of complete inactivity and we just mention it for the sake of completeness.

As we have already emphasized the necessity of an excitatory coupling, we would like to point out the relation between synaptic coupling and sleep. We have discussed the different phases of the network with respect to the coupling strength. These findings coincide with the hypothesis of Tononi that sleep is caused by an increase of synaptic strength during wakefulness \cite{tononi2003}. On the other hand, the existence of a state that shows slow oscillations when externally stimulated and otherwise does not, reflects the intuitive aspect that a tired subject with increased synaptic connections might react more sensitive on a periodic (sensory) stimulation and promotes a movement towards a deeper sleep stage.

\section{Mean field}
In order to understand the dynamics of the presented model we construct a mean field model from the time discrete system and study its fixed points and stability. For reasons of easier handling the eqs.\ (\ref{eq2a}-\ref{eq2c}) have to be slightly modified first. The currently local inhibition is replaced by a global mechanism, i.e.\ the variables $\mu$ and $\vartheta$ will depend on the whole populations of neurons, therefore they both loose their index and $x^t_i$ in the integration term of eq.\ (\ref{eq2b}) is substituted by the average network activity $x^t = (1/N) \sum_i x_i^t $. This modification obviously represents a severe abstraction of the underlying mechanism within the network, in particular with respect to the adaptive inhibitory currents assumed for each single neuron. However experimental findings reveal a strong synchrony of neurons during slow oscillation that range over large cortical areas \cite{volgushev2006}, i.e.\ for our case we assume such a high degree of synchrony among the neurons.

\begin{figure}[t]
 \centering
	\includegraphics[scale=.4]{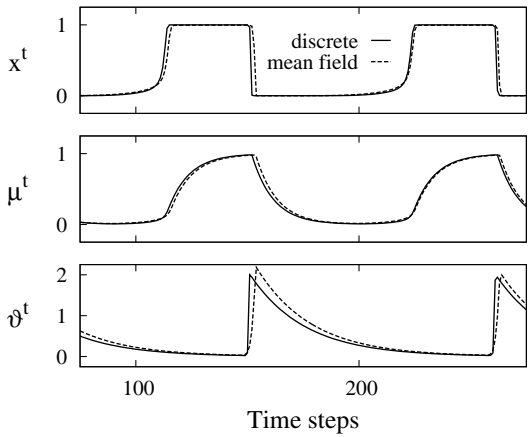}
	\caption{Comparison between the time-discrete system (solid line) and the mean field model (dashed).}
	\label{fig6}
\end{figure}

With these modifications we may rewrite the Heaviside step functions in terms of a probability, to what extent the element $x$ is active as a sigmoid function \cite{schuster2001}

\begin{equation}
 x^{t+1}\approx\Bigl(1+e^{-\beta(Cx^t-\vartheta^t-d_f)}\Bigr)^{-1}
 \label{eqMean}
\end{equation}
where $\beta$ (reminiscent of an inverse temperature) is related to the variance $\sigma$ of the Gaussian noise $\xi$ by $\beta=1/\sqrt{2\sigma^2}$. The changes yield the following mean field model
\begin{subequations}
 \begin{align}
  x^{t+1} & = \Bigl(1+e^{-\beta(Cx^t-d_f-\vartheta^t)}\Bigr)^{-1} \label{eq3a} \\
		\mu^{t+1} & = \lambda_\mu\, \mu^t+g\, x^t \label{eq3b} \\
		\vartheta^{t+1} & = \lambda_\vartheta\,\vartheta^t+h\,\Bigl(1+e^{-\beta(\mu^t-d_b)}\Bigr)^{-1}~. \label{eq3c}
 \end{align}
\label{eq3}
\end{subequations}
To legitimate the mean field model, fig.\ \ref{fig6} shows a comparison between the time discrete system with global inhibition and the mean field model. One can see clearly that both models yield the same behaviour, the discrepancy at the increase of $\vartheta$ is due to the second sigmoid function in eq. (\ref{eq3c}) that introduces a smoother transition.

If a fixed point $(x^*,\mu^*,\vartheta^*)$ is present, eqs.\ (\ref{eq3b}) and (\ref{eq3c}) can be transformed into
\begin{subequations}
 \begin{align}
  \mu^* & = \frac{g\, x^*}{1-\lambda_\mu} \label{eq4a} \\
		\vartheta^* & = \frac{h}{1-\lambda_\vartheta}~\Bigl(1+e^{-\beta(\mu^*-d_b)}\Bigr)^{-1}~, \label{eq4b}
 \end{align}
\end{subequations}
which, after insertion into eq.\ (\ref{eq3a}), leads to the (implicit) fixed point equation of the mean field model
\begin{subequations}
\begin{align}
 	x^* & = \frac{1}{1+e^{-\gamma}} \label{eq5}
\intertext{with}
	 \gamma & = \beta\biggl(C\,x^* - d_f - \frac{h}{(1-\lambda_\vartheta)(1+e^{-\delta})}\biggr) \\
		\delta & = \beta\biggl(\frac{g\,x^*}{1-\lambda_\vartheta}-d_b\biggr)~.
 \end{align}
\end{subequations}
Eq.\ (\ref{eq5}) provides the number and positions of all existing fixed points. The stability of the fixed points then follows from a linear stability analysis, where the same parameters as in the previous section were used, except $g=0.05$, $\beta=30$ and $C=1$. The adjustment of $g$ had to be performed as the mean field transformation with its two sigmoid functions causes a shift in parameter space.

\begin{figure}[t]
 \centering
 \includegraphics[scale=.4]{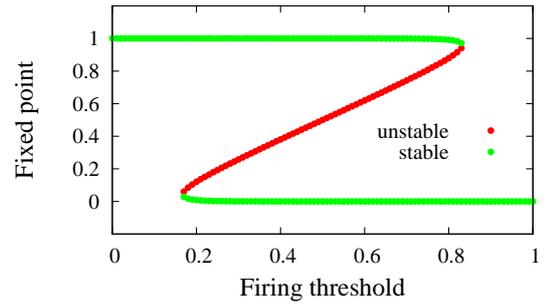}
	\caption{(Colour online) Fixed points according to eq.\ (\ref{eq5}) for a firing threshold ranging between 0 and 1. The stability is denoted by different colours: Green for stable and red for unstable fixed points. One can clearly identify an area of bistability where two stable fixed points, corresponding to ground and firing state, with an unstable fixed point in-between, coexist.}
	\label{fig7}
\end{figure}

The dynamics of the fixed points are best demonstrated by varying the firing threshold $d_f$ over a range from 0 to 1, by which we can identify three areas of stability within the model, as shown in fig.\ \ref{fig7}. 

For $d_f \lesssim 0.167$ eq.\ (\ref{eq5}) yields one fixed point, which is stable as the absolute values of all eigenvalues are smaller than 1. The firing threshold in this range is too low so that the neurons are always able to overcome $d_f$ and fire, which also explains the fixed point value of about 1.

At $d_f\approx 0.165$, a critical fixed point at $x=0.04$ emerges that commences a saddle node bifurcation, as with increasing $d_f$ this critical fixed point separates into two fixed points, of which one is stable and the other unstable. From that point on the system possesses in total two stable fixed points with one unstable in between, i.e.\ the system is bistable and either converges into the ground or firing state or, if externally stimulated, allows for switching between both.

For $d_f\approx 0.835$ another saddle node bifurcation occurs, but this time the unstable and the upper fixed points merge and vanish, leaving only the lower fixed point $x=0$. This transition corresponds to the moment when the firing threshold is too high for the neuron to overcome and it is bound to stay in the ground state.

\section{Conclusion}
We presented a simple model on a basis of a time-discrete map that generates slow oscillations. The transition from an Up to a Down state is modelled referring to an integration and inhibition process via adaptive inhibitory currents proposed by Sanchez-Vives \cite{sanchezvives2000}. An investigation of an all-to-all coupled network demonstrated above all the need of excitatory connections between neurons in order to trigger and maintain Up states.

Here we have shown that our model reproduces the experimental results from Shu \etal \cite{shu2003}, with respect to the on/off switching of network activity by external pulses of same polarity on a satisfactory qualitative level and explains them via a simple mechanism. The network possesses different states of activity: Two states where Up states are either caused by internal noise and coupling or induced through external stimulation and a third, where they are completely absent and the network shows homogeneously distributed random activity.

We conclude from our model that the coupling strength is an essential control parameter. This is roughly consistent with the hypothesis that sensory stimulation during the day strengthens synaptic connections and promotes the slow wave activity during deep sleep.

Regarding the phase portrait in fig.\ \ref{fig5} it seems obvious that another way to induce a phase transitions in our network can be realized by varying the noise intensity, i.e. a horizontal movement in the phase plane. If the blue ``active, no Up states''~phase is then labeled as the waking state, the disappearance of slow oscillations follows from an increased noise input. With this approach our model could explain brief sleep-wake transitions as they have been investigated by Lo \etal \cite{lo2002,lo2004}, who modelled complex interactions between sleep active and wake-promoting neurons by a random walk. Hence it would be worthwhile to investigate the effect of a fluctuating noise intensity close to such a phase transition.

It is undeniable that the model we presented here is a very minimal model, capturing the essential mechanisms only and allows for further extensions. We have only studied the consequences of next neighbour coupling briefly, where results were qualitatively the same. Also delay and refractory effects, which become relevant on time scales close to a single spike, were neglected as slow oscillations operate on a coarser time resolution and thus made those properties irrelevant for the modelling. In a similar way synaptic plasticity on shorter time scale may effect the dynamics as well; here there was no need to rely on plasticity in order to demonstrate the basic effects.

An interesting perspective would be to apply control and influence the dynamics. An essential part of the dynamical mechanism are the two global inhibitory degrees of freedom per neuron. Any means of control of the system is most effectively applied by taking into account the previous state of the system, which in our model would be the state of the two slow adaptive variables $\mu$ and $\vartheta$. This supports the hope that predictive, previous state-based or anticipatory chaos control techniques \cite{claussen1998,voss2003,ciszak2005,krug2007,pyragas2008} could be applied to influence the cortical Up and Down dynamics that manifest slow wave sleep.

\acknowledgments
The authors thank J. Born, L. Marshall and M. M\"olle for fruitful discussions. This work was supported by the DFG (Graduate School 235 and SFB 654-A8 ``Plasticity and Sleep'').

\end{document}